\documentclass[aps,prb,twocolumn,superscriptaddress,showpacs]{revtex4}
\usepackage{amsmath}
\usepackage{amssymb}
\usepackage{graphicx}

\begin{document}


\newcommand{\beq}{\begin{equation}}
\newcommand{\eeq}{\end{equation}}
\newcommand{\bi}{\begin{itemize}}
\newcommand{\ei}{\end{itemize}}
\newcommand{\bea}{\begin{eqnarray}}
\newcommand{\eea}{\end{eqnarray}}
\newcommand{\ba}{\begin{array}}
\newcommand{\ea}{\end{array}}
\newcommand{\bt}{\begin{tabular}}
\newcommand{\et}{\end{tabular}}
\newcommand{\bc}{\begin{center}}
\newcommand{\ec}{\end{center}}
\newcommand{\bal}{\begin{align}}
\newcommand{\eal}{\end{align}}
\newcommand{\bdisp}{\begin{displaymath}}
\newcommand{\edisp}{\end{displaymath}}
\newcommand{\ax}{\alpha}
\newcommand{\bx}{\beta}
\newcommand{\cx}{\gamma}
\newcommand{\dx}{\delta}
\newcommand{\ox}{\omega}
\newcommand{\lx}{\lambda}
\newcommand{\lb}{\bar{\lambda}}
\newcommand{\sx}{\sigma}
\newcommand{\sbar}{\bar{\sigma}}
\newcommand{\gx}{\gamma}
\newcommand{\Px}{\Phi}
\newcommand{\tx}{\theta}
\newcommand{\Pb}{\bar{\Phi}}
\newcommand{\tb}{\bar{\theta}}
\newcommand{\ab}{\bar\alpha}
\newcommand{\bb}{\bar\beta}
\newcommand{\cb}{\bar\gamma}
\newcommand{\db}{\bar\delta}
\newcommand{\Sx}{\Sigma}
\newcommand{\Lx}{\Lambda}
\newcommand{\Ox}{\Omega}
\newcommand{\Dx}{\Delta}
\newcommand{\Gx}{\Gamma}
\newcommand{\Oxb}{\bar{\Omega}}
\newcommand{\ux}{\upsilon}
\newcommand{\vtx}{\vartheta}
\newcommand{\ex}{\epsilon}
\newcommand{\vex}{\varepsilon}
\newcommand{\vphi}{\varphi}
\newcommand{\kx}{\kappa}
%
\newcommand{\cC}{\mathcal{C}}
\newcommand{\cD}{\mathcal{D}}
\newcommand{\cL}{\mathcal{L}}
\newcommand{\cLk}{\mathcal{L}_{kin}}
\newcommand{\cLm}{\mathcal{L}_m}
\newcommand{\cA}{\mathcal{A}}
\newcommand{\cV}{\mathcal{V}}
\newcommand{\RE}{\textrm{Re} \,}
\newcommand{\IM}{\textrm{Im} \,}
\newcommand{\M}{\mu}
\newcommand{\nn}{\nonumber}
\newcommand{\uu}{{\uparrow\uparrow}}
\newcommand{\dd}{{\downarrow\downarrow}}
\newcommand{\du}{{\downarrow\uparrow}}
\newcommand{\ud}{{\uparrow\downarrow}}
\newcommand{\K}{Keldysh}
\newcommand{\tm}{\widetilde{m_0}}
\newcommand{\Tr}{\mathrm{Tr}}
\newcommand{\dr}{\mathrm{d}}
\def\jnote#1{{\bf[JS: #1]}}

\title{Spin Torque Dynamics with Noise in Magnetic Nano-Systems}
\author{J. Swiebodzinski$^{1,2}$, A. Chudnovskiy$^1$,T. Dunn$^3$, and  A. Kamenev$^{3,4}$ \\
$^1$ I. Institute of Theoretical Physics, University of Hamburg, Jungiusstr. 9, 20355 Hamburg, Germany. \\
$^2$ Theoretische Physik, Universit\"at Duisburg-Essen and CeNIDE, 47048 Duisburg, Germany. \\
 $^3$Department of Physics, University of Minnesota, Minneapolis, Minnesota 55455, USA. \\
 $^4$Fine Theoretical Physics Institute, University of Minnesota, Minneapolis, Minnesota 55455, USA.}

\begin{abstract}

We investigate
the role of equilibrium and nonequilibrium noise in the magnetization dynamics on mono-domain ferromagnets.
Starting from a microscopic model we present a detailed derivation of the spin shot noise correlator.
We investigate the ramifications of the nonequilibrium noise on the spin torque dynamics, both in the
steady state precessional regime and the spin switching regime. In the latter case we apply a generalized Fokker-Planck approach to
spin switching, which models the switching by an Arrhenius law with an effective elevated temperature.
We calculate the renormalization of the effective temperature due to spin shot noise and show that the nonequilibrium noise
leads to the creation of cold and hot spot with respect to the noise intensity.

\end{abstract}

\maketitle

\section{Introduction}

The manipulation of magnetization of a ferromagnet by means of spin-polarized currents is a key issue of the state-of-the-art spintronics concepts
(for a review see Ref[\onlinecite{Ralph-Stiles}]).
In this respect the most important phenomenon
is the so-called spin-transfer-torque (STT) effect, which was predicted by Slonczewski and Berger \cite{Slon96, Berger}: A spin-polarized current may transfer angular momentum to a free ferromagnetic layer resulting in a macroscopic torque on the latter's magnetization.
In very small ferromagnets, in which the magnetization can be assumed
spatially uniform, the STT results in the rotation of the magnetization as a whole rather than in the excitation of spin waves. The STT can in particular lead to two dynamical regimes: the reversal of the free layer's magnetization or a steady state precession of the magnetization. Due to the giant magnetoresistance effect, the dynamics of magnetization is reflected in the change of resistance of the circuit containing magnetic junctions. Current induced resistance variations, identified with STT, were reported in Ref[\onlinecite{Tsoi98}]. In subsequent years both spin torque induced magnetization reversal were observed \cite{Myers99, Katine00} as well as steady state precession \cite{Tsoi00, Kiselev03, Rippard04, Krivorotov}.
Both dynamical regimes are interesting for applications: as a tempting alternative to Oersted fields in switching the magnetization in ferromagnetic memory elements on the one hand, and as clock devices used to synchronize the CPU with other logic units, on the other.
Hence a reliable description of these phenomena is very important.

On a semi-classical level, magnetization dynamics of a mono-domain ferromagnet can be well described by the Landau-Lifshitz-Gilbert Equation (LLG)
\beq
\label{LLG}
\frac{\dr {\bf m}}{\dr t}
= -\gx_0 {\bf m} \times {\bf H}_{\rm eff} + \ax_0{\bf m} \times
\frac{\dr {\bf m}}{\dr t}
+ \frac{\gx_0}{M_s \cV}{\bf m}\times
\left( {\bf m} \times {\bf I}_s \right)
\ .
\eeq
Here ${\bf m}$ is a unit vector in the free layer's  magnetization direction,
${\bf H}_{\rm eff}$ the effective magnetic field, $\cV$ the volume of the switching element, $M_s$ the absolute value of the free layer's magnetization, $\gx_0$ the gyromagnetic ratio, $\ax_0$ the Gilbert damping parameter,
and ${\bf I}_s$ the spin polarized current.
However, as the extension of associated devices are very small, effects of noise may play a significant role  and should be included into the dynamical description.
A first inclusion of noise into the LLG was given in the seminal paper of Brown \cite{Brown63}, who considered the effect of thermal fluctuations on the dynamics of a mono-domain particle by a random component ${\bf h}^R$ of the effective magnetic field entering the LLG equation (\ref{LLG}).
As a consequence of the fluctuation-dissipation theorem one finds for
the equilibrium correlator of the random field \cite{Brown63}
\beq
\label{FDT-Brown}
\langle h_i(t) h_j(t') \rangle \propto \ax_0 k_B T \dx_{ij} \dx(t-t') \ ,
\eeq
where $h_i(t)$ denotes the $i$-th Cartesian component of the random field at time $t$, $k_B$ is the Boltzmann constant and $T$ the temperature.
Since then, temperature effects on the LLG equation have been considered, both with  \cite{LiZhang2004, Cimpoesu2008, ApalkovPRB, ApalkovJMMM, Tserkovnyak05} and without the spin torque term \cite{Garcia-Palacios,  Scholz} .
The emphasis of these approaches has been on the influence of noise on switching rates, often by performing explicit numerical calculations \cite{Garcia-Palacios,  Scholz, LiZhang2004, Cimpoesu2008}.
Since the spin-torque experiments \cite{Kiselev03, Rippard04, Ralph05, Rippard06, Mistral06} are performed under
clear nonequilibrium conditions, it is natural to address, besides temperature, other sources of noise.
One possible source of nonequilibrium noise is the spin shot noise.
By analogy with the charge shot noise the quantization of the angular momentum transfer leads to spin shot noise, which manifests itself in a random Langevin force entering the equations of motion for the free magnetic layer.
It was shown by Foros {\it et al.~}\cite{Tserkovnyak05} in the context of normal metal / ferromagnet / normal metal
(NFN) structures  that the spin shot noise is the dominant contribution to magnetization noise at low temperatures. In the realistic experiments on spin torque and spin switching 
the nonequilibrium noise starts to dominate at temperatures below 
several Kelvins.  
In Ref[\onlinecite{Swie}] it was shown that inclusion of the nonequilibrium noise into the dynamical description can explain the experimentally obseved nonmonotonic dependence of the microwave power spectrum on the voltage, as well as the saturation of the spectral linewidth at
low temperatures.

In this paper we concentrate on the effect of the nonequilibrium noise on spin-switching. Without noise, the spin switching takes place when the spin current (and with it the resulting spin torque) exceeds a critical value. The critical current is determined by the strength of the magnetic anisotropy, external magnetic field and Gilbert damping, and it can be obtained from the solution of the deterministic LLG equation (\ref{LLG}). Magnetization noise opens a possibility of activated switching at currents much less than critical. Moreover, in the presence of noise the switching becomes a random process that requires a probabilistic description. The latter is based upon the solution of the Fokker-Planck (FP) equation for the probability distribution of magnetization
as derived in Ref[\onlinecite{Swie}]. A crucial step towards this description has been made in Ref[\onlinecite{ApalkovPRB}], where the authors reduced the FP equation for magnetization to the effectively one-dimensional FP equation for the probability distribution of energies and applied it to the description of activated switching processes.

A number of experiments on current induced switching have been carried out previously
\cite{
MyersRaplph2002, UrazhdinPRL, UrazhdinAPL, Fabian2003, WiesendangerScience}. From them it was found that the noise reduces the typical switching time, as one might expect.
Myers {et al.}~\cite{MyersRaplph2002} observed a broad distributions of switching currents strongly depending on
temperature, indicating a thermally activated switching process altered by the STT.
To fit the measured data they used a
Neel-Brown model \cite{Brown63, Neel} with a field- and current-dependent potential barrier height $U(H, I)$.
In this model the probability for the magnetization to switch decays exponentially with time over a characteristic relaxation time $\tau$ that obeys the Arrhenius law $\tau \sim e^{U/k_B T}$.
An implicit assumption in the Neel-Brown theory
is that magnetization dynamics is governed by a torque from
an effective magnetic field ${\bf H}_{\rm eff}$, which is derivable from the
free energy $E({\bf M})$ of the system via ${\bf H}_{\rm eff} = -\frac{1}{\mu_0} {\bf \nabla}_M E({\bf M})$.
The spin torque however is non-conservative and the concept of a corresponding potential barrier is ill-defined, which makes the situation significantly more complicated. For thermally activated switching in presence of STT
Urazhdin {\it et al.~}\cite{UrazhdinPRL, UrazhdinAPL} found that the activation energy strongly depends on the magnitude as well as the direction of
the current. To capture the observed features they introduced an effective temperature distinct from the real temperature in the Neel-Brown formula. Its current directional dependence indicated that the heating is not the ordinary Joule heating.
Based on a stationary solution of the Fokker-Planck equation Apalkov and Visscher \cite{ApalkovPRB, ApalkovJMMM}, and Li and Zhang \cite{LiZhang2004},
linked this effective temperature to the spin torque.
In their model the alteration of switching rates is due to the change of the elevated effective temperature in the Arrhenius factor, which in general yields a non-Boltzmann probability distribution.
We extend the approach of Ref[\onlinecite{ApalkovPRB}] considering the influence of the noise on the switching in the whole range of currents, from the noise induced activated switching at small currents up to the almost deterministic switching  by large critical currents. We also take into account specific  angular dependence of the nonequilibrium noise and analyze the applicability of the effective temperature description to the nonequilibtium noise in detail.

With the present paper we hope to give a contribution towards a better understanding of noise in magnetic systems.
In section \ref{sec-Langevin} we start our discussion with a detailed derivation of the spin shot noise correlator by means of the Keldysh technique.
It is our goal to depict the underlying mechanisms that lead to the occurrence of spin shot noise in magnetic
nanodevices and to provide a general mathematical framework for their description.
We then turn our attention to the ramifications of the
noise on the magnetization dynamics.
We address
the question of switching rates estimation
in sections \ref{sec-FP} and \ref{sec-switch} by applying a generalized Fokker-Planck
approach.
Within this approach the alteration of switching rates
due to spin torque
is determined by an
effective temperature $T_{\rm eff}$. The latter differs from
the real temperature $T$, as it incorporates the effects
of the damping, the spin torque, and noise.
We calculate the renormalization of the effective temperature
due to the nonequilibrium noise.
A conclusion of our findings is given in section \ref{sec-concl}.

\section{Spin Shot Noise Correlator}
\label{sec-Langevin}

In this section, starting from a microscopic model of a magnetic tunnel junction (MTJ)
we will derive a stochastic version of the LLG equation.
Fluctuations will naturally come about
due to the nonequilibrium situation, and will comprise the random part of the stochastic LLG.
In particular, performing a perturbative expansion of the Keldysh action in terms
of the spin flip processes and the tunneling amplitude we will
be able to derive the spin shot noise correlator.

The model MTJ consists of two itinerant ferromagnets separated by a tunneling barrier.
Let us introduce the corresponding model Hamiltonian allowing for
an external magnetic field ${\bf H}$, tunneling of itinerant
electrons through the barrier and exchange coupling between
the itinerant electrons and the free layer's magnetization. It reads
\begin{align}
\label{Ham1}
    H_0 = &\sum_{k,\sigma} \epsilon_{k\sigma} c^{\dagger}_{k\sigma} c_{k\sigma} +
    \sum_{l\sigma} \epsilon_{l} d^\dagger_{l\sigma} d_{l\sigma} -\gamma  {\bf S}\cdot {\bf H}
- 2 J\, {\bf S}\cdot {\bf s} \nn \\
&
+
    \left[
    \sum_{kl\sx}
W_{kl}c^\dagger_{k\sx} d_{l\sx} + h.c.
\right] \, .
\end{align}
The notation is as follows:
The creation (annihilation) operators $c^\dag_{k\sx}$ ($c_{k\sx}$)
and $d^\dag_{l\sx}$ ($d_{l\sx}$)
describe the itinerant electrons of the fixed and the free
magnetic layer respectively.
$\sx = +$ corresponds to the respective majority
and $\sx = -$ to the minority spin band, and the
indices $k$ and $l$ label momentum.
The operator ${\bf S}$ describes the total spin of the free layer. It is connected to the free layer's magnetization via ${\bf S} = {\bf M}{\cal V}/\gamma$.  ${\bf s}=\frac{1}{2}\sum_{l\sigma\sigma'} d^\dagger_{l\sigma} \vec\sigma_{\sigma\sigma'} d_{l\sigma'}$ is the quantum operator associated with the spin of itinerant electrons, where $\vec\sigma$ denotes the vector of Pauli matrices. $J$ is the exchange coupling constant and $W_{kl}$
are tunneling matrix elements.

For the subsequent discussion we
assume that the time between two tunneling processes is much larger
than the relaxation time in the free ferromagnet, which is equivalent to
assuming
a complete spin relaxation in the free magnetic layer.
This allows us to introduce an
\emph{instantaneous} reference frame with the spin quantization axis
directed along the free layer's magnetization direction.
To render the free layer's magnetization a dynamical variable,
we make use of the Holstein-Primakoff parametrization \cite{Holstein-Primakoff}
\begin{equation}
S_z=S- b^\dagger b;   \  \
S_-= b^\dagger\sqrt{2S-b^\dagger b};   \  \
S_+= \sqrt{2S-b^\dagger b} \,b,
 \label{HP-rep}
\end{equation}
where $b^\dagger, b$ are usual bosonic operators and
$S_\pm = S_x \pm i S_y$. At low temperatures we can assume
that the expectation value of $b^\dag b$ is much smaller
than $2S$ allowing to treat the square root to zeroth order in $b^\dag b$.
Taking all of the above mentioned into account,
Hamiltonian (\ref{Ham1})
can be written in the instantaneous reference frame as
\begin{eqnarray}
\label{Ham3}
\nonumber
&&    H_0=\sum_{k,\sigma} \epsilon_{k\sigma} c^{\dagger}_{k\sigma} c_{k\sigma} +
    \sum_{l\sigma}(\epsilon_{l} -  J S\sigma) d^\dagger_{l\sigma} d_{l\sigma}
  -\gamma  S H_z \nn\\
&&   + \gx b^\dag b H_z+ Jb^\dag b\sum_{l\sigma}  \sigma  d^\dagger_{l\sigma} d_{l\sigma}
  \nn +
    \left[
    \sum_{kl,\sigma\sigma'}
W_{kl}^{\sigma\sigma'} c^\dagger_{k\sigma} d_{l\sigma'} \right.  \\
&&   \left.
- \, b\sqrt{2S}\left(J\sum_l d^\dag_{l\downarrow}d_{l\uparrow} + \frac{\gx}{2} H_-\right) + h.c.
\right]
\, ,
\end{eqnarray}
where we used the notation $H_\pm = H_x \pm i H_y$.
$W_{kl}^{\sx\sx'} $ are spin dependent tunneling
matrix elements given by
\begin{align}
& W_{kl}^{\sx\sx'} = \langle \sx | \sx' \rangle W_{kl} ,\\
 \quad
& \langle \sx | \sx \rangle = \cos\tfrac{\tx}{2}e^{-\frac{i}{2}\sx\phi} , \quad
\langle \sx | \sx' \rangle = \sx'\sin\tfrac{\tx}{2}e^{\frac{i}{2}\sx\phi} \ .
\end{align}

Hamiltonian (\ref{Ham3}) can be now readily translated into a Keldysh
action using the general scheme of the Keldysh technique \cite{Kamenev05}.
To this end we switch to
symmetric (''$cl$") and antisymmetric (''$q$") linear combinations
of the field operators. In accordance with parametrization
(\ref{HP-rep}) the former are connected to the
$m_\pm$ components of the
free layer's magnetization in the instantaneous
reference frame via
\beq
b_{cl}(t) = \sqrt{\frac{M_s \cV}{2\gx}}\, m_+ (t) \ , \qquad
\bar b_{cl}(t) = \sqrt{\frac{M_s \cV}{2\gx}}\, m_- (t) \ .
\eeq
For the retarded and advanced components of the fermionic Green functions for
the itinerant electrons of the free and fixed layer we obtain in the energy domain
\beq
G^{R/A}_{l\sx}
=
\frac{1}{\ex - \ex_{l\sx} \pm i0}
 \ , \qquad  G^{R/A}_{k\sx} = \frac{1}{\ex - \ex_{k\sx} \pm i0}  \ ,
\eeq
where $\ex_{l\sx}= \ex_{l} - \sx JS$ are the energies of
the itinerant electrons with momentum $l$ and spin $\sx$
in the free ferromagnet, and $\ex_{k\sx}$ the corresponding
energies for the fixed layer. The Keldysh components are
\begin{align}
G^{K}_{l\sx} = (1-2n^d_F(\vex))\dx \left(\ex - \ex_{l\sx} \right), \\
 G^{K}_{k\sx} = (1-2n^c_F(\vex))\dx \left(\ex - \ex_{k\sx} \right)  ,
\end{align}
where chemical potentials $\mu_{d/c}$ for the free and fixed layer are included in the fermionic distribution functions $n_F^{c/d}$.
For future use we also define the matrices in Keldysh space
\beq
\label{gx-mat}
\gx^{cl} =
\begin{pmatrix}
1 & 0\\
0 & 1
\end{pmatrix} \ ,
\qquad
\gx^{q} =
\begin{pmatrix}
0 & 1\\
1 & 0
\end{pmatrix} \ .
\eeq

\begin{figure}
\begin{center}
\includegraphics[width=\linewidth]{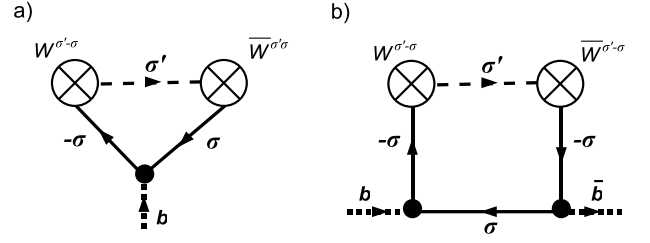}
\caption{(Color Online) Diagrams for spin flip processes: a) First order. b) Second order. Solid (dashed) lines denote electronic propagators in the free (fixed) layer. Bold dashed lines are propagators of HP bosons. Tunneling vertices are denoted by circles with crosses.}
\label{Fig-diagr}
\end{center}
\end{figure}

The Keldysh action can be now solved perturbatively in terms of the tunneling amplitude
and the spin flip processes. In second order in both quantities this leads
to the diagrams of Fig.~\ref{Fig-diagr}.
The corresponding equations of motion are obtained
when varying the action $\cA$ with respect to the quantum component
\beq
\label{eqmo-allg}
\frac{\dx \cA}{\dx b_q} = 0 \ , \qquad \frac{\dx \cA}{\dx \bar b_q} = 0 \ .
\eeq
Finally we note that, in the
instantaneous reference frame,  we have for $m_\pm = m_x \pm i m_y$
\beq
\langle m_\pm \rangle = 0 \ ,
\qquad \langle \partial_t m_\pm \rangle \neq 0 \ ,
\eeq
where, on the other hand
\beq
\langle m_z \rangle = 1 \ ,
\qquad \langle \partial_t m_z \rangle = 0 \ .
\eeq

We may now translate diagrams \ref{Fig-diagr} (a) and (b)
into the analytical expressions.
However,
let us start
with the contribution of zeroth order
(in spin flips \emph{and} in tunneling). It reads
\beq
\label{A0}
\cA_0 = \int \dr t \bar b_q (t)
\left( i \partial_t b_{cl}(t) + \gx \sqrt{S/2} H_+ \right)
+ \rm{ c.~c.}
\eeq
The resulting equations of motion are
\begin{align}
\label{eqmo-kin}
i\partial_t b_{cl} + \gx \sqrt{S/2} H_+ = 0
\end{align}
and a corresponding complex conjugate equation for $\bar b_{cl}$. Equation (\ref{eqmo-kin}) describes the precession of the magnetization around the
magnetic field ${\bf H}$ and forms the first term
of the LLG equation (\ref{LLG}).

Let us come to the diagram of Fig.~\ref{Fig-diagr}(a).
To extract its contribution to the action we have
to calculate
\beq
-J\sqrt{S}\sum_{kl\sx\sx'}W_{kl}^{\sx'-\sx}\bar W_{kl}^{\sx'\sx}
b_\sx
\Tr \left\{
G^d_{l\sx}\gx^q G^d_{l-\sx}G^c_{k\sx'}
\right\} \ ,
\eeq
where for brevity the symbolic notation $b_\sx$ with $b_\uparrow = b_q$ and $b_\downarrow = \bar b_q$ was introduced.
The resulting action reads
\beq
\label{A1}
\cA_1 = \frac{i}{2\sqrt{S}} I_s \int \dr t  \left\{ \bar b_q (t) \sin \tx e^{-i \phi} - b_q (t) \sin \tx e^{i \phi} \right\} \ .
\eeq
Variation of (\ref{A1}) with respect to $b_q$ and $\bar b_q$ gives the following contribution to the equations of motion
\begin{align}
\label{A1-var}
\frac{\dx \cA_1}{\dx b_q(t)} & = -i \frac{I_s}{\sqrt{2S}} \sin
\tx e^{i\phi} \ , \qquad
\frac{\dx \cA_1}{\dx \bar b_q(t)}  = i \frac{I_s}{\sqrt{2S}}
\sin \tx e^{-i\phi}  \ .
\end{align}
Again, using the HP parametrization (\ref{HP-rep}) and
the relation between
${\bf S}$
and
${\bf m}$,
equation (\ref{A1-var}) can be readily translated into the corresponding
equation of motion for the magnetization.
The result is the spin torque term of equation (\ref{LLG})
\beq
\partial_t {\bf m} = \frac{\gx}{M_s\cV} {\bf m}
\times \left( {\bf I}_s \times {\bf m}\right) \ .
\eeq

As far as the remaining diagram (Fig.~\ref{Fig-diagr}(b))
is concerned we have to distinguish two contributions:
One with two
quantum components and one with a quantum and a classical
component
respectively.\footnote{The $cl$-$cl$ component
vanishes by virtue of the fundamental properties of
the Keldysh formalism.}
In the first case we obtain
\begin{align}
\nonumber
 J^2 S b_q \bar b_q \sum_{kl\sx\sx'} |W^{\sx'-\sx}_{kl}|^2
 \Tr \left\{  G^d_{l-\sx}(\vex)\gx^q G^d_{l\sx}(\vex-\omega) \times
\right.  & \\
 \left.  \gx^q G^d_{l-\sx}(\vex) G^c_{k\sx'}(\vex) \right\} \ . &
\label{eq-qq-b}
\end{align}
In the second case we have
\begin{align}
\nonumber
J^2 S b_{cl} \bar b_q \sum_{kl\sx\sx'} |W^{\sx'-\sx}_{kl}|^2
\Tr \left\{  G^d_{l-\sx}(\vex)\gx^q G^d_{l\sx}(\vex-\omega)\times \right. & \\
\left.\gx^{cl}  G^d_{l-\sx}(\vex) G^c_{k\sx'}(\vex) \right\} \ , &
\label{diagrb-diss}
\end{align}
and the corresponding contribution with $q \leftrightarrow cl$. The resulting action is
\beq
\label{A2}
\cA_{2} =  \int \dr t
\left[
\bar \ax (\tx) \left(\bar b_q
\partial_t b_{cl} - \bar b_{cl}\partial_t b_q \right)
+ \frac{2i}{S}
\cD(\tx) \bar b_q b_q \right]  \ ,
\eeq
where
\begin{align}
\label{alpha-ren}
\bar \alpha(\tx) & = \frac{\hbar \gx}{e M \cV}\left(
\frac{\dr I_{sf}(\tx)}{\dr V}\right) \ , \\
\label{DI}
\mathcal{D}(\tx) & = \frac{M_s \mathcal{V}}{\gx}\ax_0 k_B T
+
\frac{\hbar}{2}\, {\rm I}_{sf}(\theta) \coth\left(\frac{eV}{2k_BT}\right)\, .
\end{align}
The spin flip current ${ I}_{sf}$ can be calculated from the
electric conductances $G_{P(AP)}$ in the parallel
(antiparallel) configuration as follows
\beq
\label{Isf}
\frac{\dr I_{sf}(\tx)}{\dr V} = \frac{\hbar}{4e}\left[ G_P \sin^2\left( \frac{\tx}{2}\right)+ G_{AP} \cos^2\left( \frac{\tx}{2}\right)\right] \ .
\eeq
Action (\ref{A2}) consists of two parts.
The first term is a damping term. In the LLG equation it
will result in a renormalization of the Gilbert damping parameter.
The renormalization is
due to the coupling to the reservoirs. The enhancement of the
damping, Eq.~(\ref{alpha-ren}), is closely related to
the spin-pumping enhanced damping
as discussed in Ref[\onlinecite{Tserkovnyak02, Halperin-Tserkovnyak}]
in the framework of the Landauer-B\"uttiker formalism.
As far as the second term of Eq.(\ref{A2}) is concerned
we introduce
a Hubbard-Stratonovich auxiliary field which decouples the action.
Let us denote this (complex) field by ${I}^R_+ =
I^R_{s, x} + i I^R_{s, y}$.
We can write
\begin{multline}
\int \dr I^R_+ \dr \bar I^R_+ \, e^{-\frac{1}{4\cD}I^R_+ \bar I^R_+}\, e^{i\cA_{21}} \\ =
\int \dr I^R_+ \dr \bar I^R_+ \, e^{-\frac{1}{4\cD}I^R_+ \bar I^R_+}\, \exp\left\{-i\tfrac{1}{\sqrt{2S}}(I^R_+ \bar b_q + \bar I^R_+ b_q ) \right\}
\ ,
\end{multline}
where we abbreviated the second term of (\ref{A2}) by $\cA_{22}$.
As one can see the result is a noise-averaged term which is linear in
the quantum component.
The linear action constitutes a resolution
of functional $\dx$-functions of the Langevin equations on
$b_{cl}(t)$ and its complex
conjugate. The stochastic properties are encoded
in the auxiliary field ${I}^R_+ $, precisely
in the correlator (\ref{DI}). For $b_{cl}$ the Langevin equations
 read
$i\partial_t b_{cl} = \tfrac{1}{\sqrt{2S}}I^R_+$. This corresponds to $i\partial_t m_{+} = \tfrac{\gx}{M\cV}I^R_+$
leading to the random term of the stochastic LLG equation.
Adopting the notation ${\bf I}^R_s \equiv \dx {\bf I}_s$,
in conclusion we have found
\beq
\partial_t {\bf m} \,   = \frac{\gx}{M_s \cV}
{\bf m} \times \left( \dx{\bf I}_s \times {\bf m}\right) \ ,
\eeq
where the stochastic field is characterized by
\beq
\label{corr}
\langle \dx{I}_{s,i}(t) \dx{I}_{s,j}(t) \rangle
= 2 \cD (\tx) \dx_{ij} \dx (t-t') \
\eeq
with the correlator $\cD(\tx)$ given by Eq.~(\ref{DI}).

To complete our discussion we add some comments concerning
the correlator (\ref{DI}).
To start with, we note that $\cD$ contains two parts,
an equilibrium part (which is phenomenological, and in compliance
with the FDT proportional to $\ax_0$ taking into account
intrinsic damping processes)
and a nonequilibrium part. The nonequilibrium part
exhibits a dependence on the mutual orientation
of the fixed and free layer's magnetizations.
This angle dependence enters
the
correlator through the spin flip current $I_{sf}$.
The physical meaning behind this quantity is the following:
$I_{sf}$  counts the total number of spin flip events,
irrespective of their direction. Hence, even if there is no
contribution to the spin current $I_s$, the spin flip current
$I_{sf}$ may acquire a nonzero value.
The discreteness of angular momentum transfer in each spin flip
event leads to the occurrence of the nonequilibrium noise.
In this sense
the nonequilibrium part of (\ref{DI})
can be identified with the
spin
shot noise.

In conclusion we have derived the following stochastic LLG equation
\beq
\label{LLGstat}
\frac{\dr {\bf m}}{\dr t}
= -\gx_0 {\bf m} \times {\bf H}_{\rm eff} + \ax_0{\bf m} \times
\frac{\dr {\bf m}}{\dr t}
+ \frac{\gx_0}{M_s \cV}{\bf m}\times
 \left[ {\bf m} \times \left( {\bf I}_s + {\bf I}^R_s\right) \right]
\ ,
\eeq
where the random field correlator is given by Eq.~(\ref{corr}) along with (\ref{DI}) and (\ref{Isf}) .

\section{Fokker-Planck Approach to Spin Torque Switching}
\label{sec-FP}

Spin torque switching is observable in
two different regimes.
On the one hand, the spin torque
can switch the magnetization of a free ferromagnet
when the
current exceeds a critical value $I_c$. On the other hand,
switching is also observed for currents below $I_c$.
In the second case the actual switching procedure
is mainly noise induced.
A suitable description of switching times
in this regime can be obtained from the
Fokker-Planck approach which was recently introduced
by Apalkov and Visscher in the context of thermal
fluctuations \cite{ApalkovPRB, ApalkovJMMM}.
Within this approach switching rates are specified by
an Arrhenius like law with an \emph{effective} temperature
$T_{\rm eff}$. The latter differs from the real temperature
$T$, as it is influenced by the damping and the spin torque.
In the sequel we present a generalization of the
method to nonequilibrium noise, and
show that the spin shot noise alters the effective temperature.

Let us start our consideration with the Fokker-Planck equation
as introduced by Brown \cite{Brown63}.
We denote the probability density
for the magnetization of a mono-domain particle
by
$\rho({\bf m}, t)$. The corresponding Fokker-Planck equation
can be written
in the form of a continuity equation
\beq
\label{fp-brown}
\frac{\partial \rho({\bf m, t})}{\partial t} = - {\bf \nabla} \cdot {\bf j}({\bf m}, t)
\eeq
with probability current \cite{Brown63}
\beq
\label{j}
{\bf j}({\bf m}, t) = \rho({\bf m}, t){\bf \dot m}_{det}({\bf m}) - D {\bf \nabla}\rho({\bf m}, t) \ .
\eeq
Here ${\bf \dot m}_{det}$ denotes the deterministic part
of the stochastic LLG (\ref{LLG}) and $D$ is the random field
correlator.
We recall that the dynamics governed by (\ref{LLG})
conserves the absolute value of ${\bf m}$.
As a consequence the movement of the
tip of ${\bf m}$ is
restricted to the surface of a sphere, which we will
call the $m$-sphere. The gradient and the divergence
in (\ref{fp-brown}) and (\ref{j}) are  2-dimensional objects,
both living on the $m$-sphere.

We now observe that in presence of anisotropy
the phase space will be in general separated.
The potential landscape will exhibit different minima referring to
stable and meta-stable states of the magnetization. Precession
of the magnetization
takes place around one (or more) of these equilibrium
positions. We refer to orbits of constant energy as
Stoner-Wohlfarth (SW) orbits.
Now,
considering the dynamics of the magnetization vector one
can distinguish two different time scales.
The time scale
for the angular movement, on the one hand,
is characterized by
the precession frequency.
On the other hand there is also a time
scale for a possible change in energy.
In the following we will require that
the time scale for the
change
in energy is much longer
than the time scale for constant energy precession.
In other words: we assume
that the magnetization vector stays rather long on a
SW orbit before changing to higher/lower
energies.
In this low damping and small current limit we can
introduce an energy-dependent probability density
by identifying $\rho'_i(E({\bf m}), t) \equiv \rho ({\bf m}, t)$, where the index $i$ takes into account that the energy dependence may be different in different regions of the $m$-sphere.
The above mentioned time scale separation allows  us
to average out the movement along the
 SW orbit
and to be concerned
with only the
long time dynamics.

The idea of the FP approach is to translate
equation (\ref{fp-brown}) into a corresponding
equation for $\rho'_i(E)$.
For thermal noise this has been done
in Ref[\onlinecite{ApalkovPRB}].
We now give a generalization of the method
to the angle-dependent spin shot noise
of section \ref{sec-Langevin}. To this end we write the
correlator (\ref{DI}) in the form
\beq
D(\tx) = D_{th} + D_0 \left[ 1 - P \cos\tx \right] \ ,
\eeq
where $D_{th} = \frac{\gamma \alpha k_B T}{M_s \mathcal{V}}$ is the thermal part and $D_0[1-P\cos\theta]$ the nonequilibrium part of the correlator. \footnote{Note that we use here a Langevin term in form of an effective field rather than the current, resulting in a different prefactor as compared to Eq.~(\ref{DI}).}

We abbreviated the angle-independent part of the spin shot noise by $D_0$.
We also used
\beq
P = \frac{G_P - G_{AP}}{G_P + G_{AP}} \ .
\eeq

In general we can write the
Fokker-Planck equation for the
distribution $\rho'_i(E({\bf m}), t)$
in the form \cite{ApalkovPRB}
\beq
\label{FP-E-Apalk}
\frac{\gx P_i(E)}{M_s\mu_0} \frac{\partial \rho_i'(E, t)}{\partial t}
= - \frac{\partial}{\partial E} j^E_i (E, t) \ ,
\eeq
where $P_i(E)$ is the period of the orbit with energy $E$. $j^E_i$ is the probability current in energy.
It is given by
\begin{align}
\label{jE}
& j^E_i(E, t) =
\oint [{\bf j}({\bf m}, t) \times d {\bf m}] \cdot {\bf m} =
-\gx \ax \rho'_i(E, t)I^E_i(E) \nn \\
& + \gx J \rho'_i(E, t){\bf m}_p \cdot {\bf I}^M_i  - \frac{\partial \rho'(E)}{\partial E} M_s D_{th}I^E_{\tx, i}  \ .
\end{align}
The constant $J$ is defined in such a way that $J{\bf m}_p = \gx/(M \cV){\bf I}_s$ if ${\bf m}_p$ is a unit vector in direction of ${\bf I}_s$.
Furthermore we have introduced the following integrals along the
SW orbit
\begin{align}
\label{IE-th}
I^E_{\tx, i} &= I^E_i +
\frac{D_0}{D_{th}}
\left(I^E_i - P\oint \cos\tx H_{eff} \dr m \right) \ , \\
\label{IE}
    I^E_i(E) &= \oint H_{eff} \dr m \\
\label{IM}
{\bf I}^M_i(E, t) & = \oint \dr {\bf m} \times {\bf m} \ .
\end{align}
A steady state solution of the FP equation is obtained by setting $j^E_i = 0$. From (\ref{jE}) we get the following differential equation for the probability density $\rho'_i$:
\beq
\label{beta-neq}
\frac{\partial \ln \rho'_i(E)}{\partial E} = \frac{\gx}{D_{th}M_s}\lx_i(E)[-\ax + \eta_i(E)J] \equiv - \mathcal{V}\bx'_i(E) \ ,
\eeq
where the right hand side serves as a definition of an inverse effective temperature $\bx'_i(E)$. From (\ref{beta-neq}) one can see that, depending on the sign of the spin current, the spin torque may either enhance or diminish the damping, leading to a lower or higher effective temperature, respectively.
In (\ref{beta-neq})
we have defined
\beq
\label{eta}
\eta_i(E) = \frac{{\bf m}_p \cdot {\bf I}^M_i(E)}{I^E_i(E)} \
\eeq
and
\beq
\label{lx}
\lx_i(E) = \frac{I^E_i}{I^E_{\tx,i}}
\ .
\eeq
$\eta_i$ can be viewed of as the ratio of the work of the Slonczewski torque to that of the damping \cite{ApalkovPRB}.
The quantity $\lx$ gives the renormalization of the effective
temperature as compared to the pure thermal case.
We can write
for $\lx$
\beq
\label{lxT}
\lx(E)
= \frac{T_{\rm eff}}{T_{\rm eff}'}
\ ,
\eeq
where $T_{\rm eff}$ is the effective temperature when only
equilibrium noise is present, and $T_{\rm eff}'$ the
effective temperature when both, equilibrium and nonequilibrium
noise are included.
It should be observed from (\ref{beta-neq}) that
the effective temperature is in general energy dependent.
The corresponding probability distribution will thus, in general,
differ from the
Boltzmann distribution.
However, when we turn off the nonequilibrium, $\lx(E)\equiv 1$
and $J=0$. In this case the solution of (\ref{beta-neq})
is exactly a Boltzmann distribution.

In the remainder of this section we evaluate $\lx$ for
an exemplary system with easy axis and easy plane an\-isot\-ropy.
The easy axis is chosen to be the $z$ axis and the easy
plane is the $y$-$z$ plane.
The magnetization direction of the
fixed layer, ${\bf m_p}$,
is taken to be anti-parallel to the $z$ axis.
Let us use the following convention for the spherical coordinates:
$m_x = \cos \vtx$, $m_y = \sin \vtx \sin \vphi$,
$m_z = \sin \vtx \cos \vphi$.
The SW condition defines the orbits of constant energy. For our system it reads
\beq
\label{sw-1}
\tfrac{E({\textbf{M}})}{\mu_0} = -\tfrac12 H_K M_S ({\textbf m e_z})^2
+\tfrac12 M_S^2 ({\textbf m e_x})^2 \ .
\eeq
We abbreviate $\kappa_1 = H_K M_S$ (characterizing the strength of easy-axis an\-isot\-ropy), $\kx_2 = M_S^2$ (characterizing the strength of easy-plane an\-isot\-ropy) and $d = \frac{\kx_1}{\kx_2}$, being the ratio of easy-axis to easy-plane anisotropy,
so that (taking the magnetic constant $\mu_0 = 1$) we can obtain from equation (\ref{sw-1}) the dimensionless energy $c$
\beq
\label{sw-bahn}
c \equiv \tfrac{2 E}{\kx_2} = - d m_z^2 + m_x^2 \ .
\eeq
This relation defines the 'potential landscape' of our system.
We can distinguish
three regions: Two potential wells, one around $\vphi=0$ (well 1)
and one around $\vphi=\pi$ (well 2), and a third region (region 3)
with energies above the saddle point energy, separating the two
wells. Switching takes place if the magnetization vector
changes from some orbit in the one well to an orbit in the other well.
Equation (\ref{sw-bahn}) defines the orbits of integration
for the evaluation of (\ref{lx}). Let us concentrate on
orbits lying in the potential well around $\varphi=0$
with energies $c \leq 0$.\footnote{Note that due
to the symmetry of (\ref{sw-1}) this can be done
without loss of generality.}
In addition we assume a
strong easy plane anisotropy, allowing to consider
small deviations of $\vtx$ around $\frac{\pi}{2}$.

\begin{figure}
\begin{center}
\includegraphics[width=0.8\linewidth]{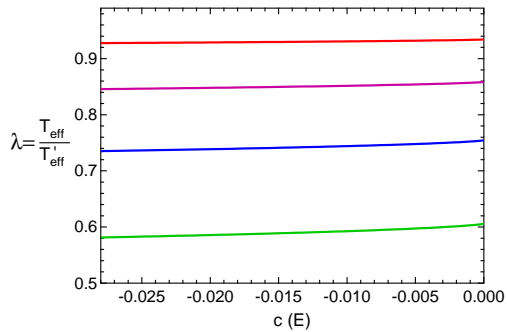}
\caption{(Color Online) $\lx=T_{eff}/T'_{eff}$ as a function of $c$ in the case ${\bf m}_p \uparrow \downarrow {\bf e}_z$ for $eV=k_BT$ (red), $eV=5 k_BT$ (magenta), $eV=10k_BT$ (blue), $eV=20k_BT$ (green).}
\label{Fig-ges-ap}
\end{center}
\end{figure}
We
fix the Gilbert damping to $\ax=0.01$, the ratio of anisotropies to $d=0.028$, the polarization to $P=0.81$, and $M_s \mathcal{V}/\gx=10\hbar$. These values define the ratio $D_0/D_{th}$ as a function of $eV/k_B T$. The results for $\lx={T_{eff}}/{T'_{eff}}$ are plotted  in Fig.~\ref{Fig-ges-ap}.
As can be seen
from
the plot
taking into account
the nonequilibrium noise results in a renormalization
of the effective temperature. This renormalization
is proportional to the applied voltage $V$ and can be very
strong for sufficiently large values of $V$.
The deviation from the purely thermal case ($\lx=1$) approaches
15\%
for $eV=5k_BT$ and is thus experimentally not negligible! For $eV=10k_BT$
the deviation is even in the order of 25\% and grows further with the voltage.
The variation
of $\lx$ with energy is
on the other hand very weak.
This indicates that the influence of the angle dependence is
rather small or in other words:
The angle-dependence of the correlator does not lead to
a significant variation of $T_{eff}$ with precession orbit.

Let us continue our discussion of the renormalized effective
temperature by considering the limit where
the equilibrium part
of the correlator is much smaller than
its nonequilibrium part
and thus may be neglected.
In this case we define the following quantity
of interest
\beq
\label{lx'/D0}
{\lx'(E)} =
  \frac{D_0}{D_{th}} \lx(E) \ .
\eeq
One should note the difference between
$\lx$ and $\lx'$. From (\ref{lxT}) we see that
$\lx$ is the ratio of the effective temperatures $T_{eff}$ and $T'_{eff}$
for systems without and with nonequilibrium noise respectively.
On the other, from the
definition (\ref{lx'/D0}) it is clear that $\lx'$ is a measure for the
influence of the angle dependence of the correlator. The stronger $\lx'$
deviates from $\lx'=1$ the stronger is the influence of the angle-dependence.

\begin{figure}
\begin{center}
\includegraphics[width=0.8\linewidth]{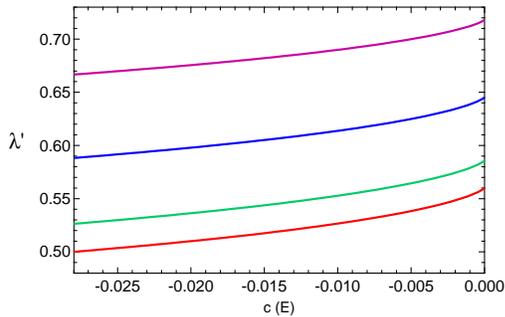}
\caption{ (Color Online)
$\lx'$ as a function of $c$ in the case ${\bf m}_p \uparrow
\downarrow {\bf e}_z$ for $P=1$ (red), $P=0.9$ (green),
$P=0.7$ (blue), $P=0.5$ (magenta). $d=0.028$.
}
\label{Fig3-apz}
\end{center}
\end{figure}

In Fig.~\ref{Fig3-apz} we plot $\lx'$
for our model system for $d=0.028$ and different values of $P$.
As one can see from Fig.~\ref{Fig3-apz} the largest deviation from
$\lx'(E)=1$ (corresponding to the strongest influence of the
angle-dependence) is present
at the minimum of the well ($c=-d=-0.028$).
The smallest deviation from $\lx'(E)=1$ is observed for
orbits which lie near the separatrice. The overall change
of $\lx'(E)$ for $P=1$ is of the order of 10\%.

\begin{figure}
\begin{center}
\includegraphics[width=0.9\linewidth]{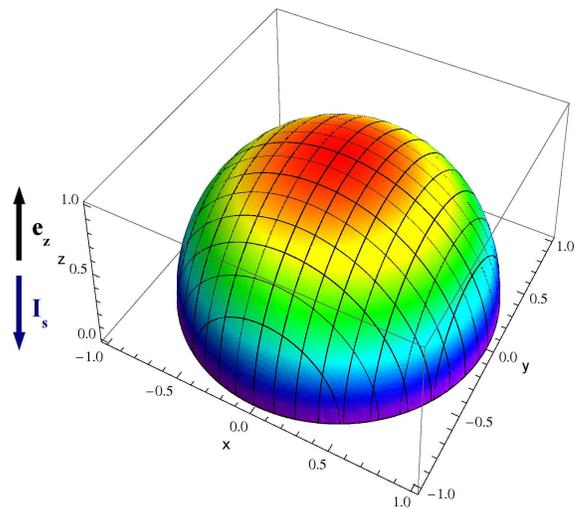}
\caption{(Color Online) Hot spots (red) and cold spots (blue) on the $M$-half-sphere in case of ${\bf m}_p \uparrow \downarrow {\bf e}_z$. The noise intensity is highest at the bottom of the well.}
\label{Fig-cold-spots}
\end{center}
\end{figure}

These results provide a good insight
into the influence of the angle-dependence.
As $\lx' \sim 1/T'_{eff}$, a small value
of $\lx'$ indicates a "hot" spot whereas
large values of $\lx'$ correspond
to "cold" spots on the $m$-sphere.
For the particular system under
consideration, \emph{cf.}~Eq.~(\ref{sw-1}),
the equilibrium position of the magnetization
is roughly along the $z$ axis. SW orbits of
precession are
symmetric with respect to this axis. At the bottom of the well
$\tx = \pi$ and the spin shot noise has its maximal value.
We hence expect a hot spot at the minimum of the well.
With increasing energy the orbits will become larger. The
angle $\tx$ will vary along these orbits. However
as the orbit energy grows
the trajectories increasingly go through regions
of smaller $\tx$, so that the average value of $\tx$
will diminish with orbit energy.
As a consequence
the nonequilibrium noise will become smaller as well.
Cold orbits should be hence those that are in the vicinity of
the separatrice. This is exactly what can be read off
from Fig.~\ref{Fig3-apz}. Our findings are thus in agreement with the geometrical situation.
Cold spots and hot spots on the $m$-sphere are shown in
Fig.~\ref{Fig-cold-spots}.

\section{Switching time of spin-torque structures}
\label{sec-switch}

The switching process can be analyzed by performing numerical simulations
of the Langevin  equations of motion with the inclusion of temperature and shot noise
via the random field  term. In this section we present such simulations
for  Gilbert damping of $\alpha=0.01$, an anisotropy ratio of
$d=0.028$, and with a spin torque current characterized by $J$ and
polarized in the $\mathbf{m}_{p}=-e_{z}$ direction.

\begin{figure}
\begin{centering}
\includegraphics[width=8cm]{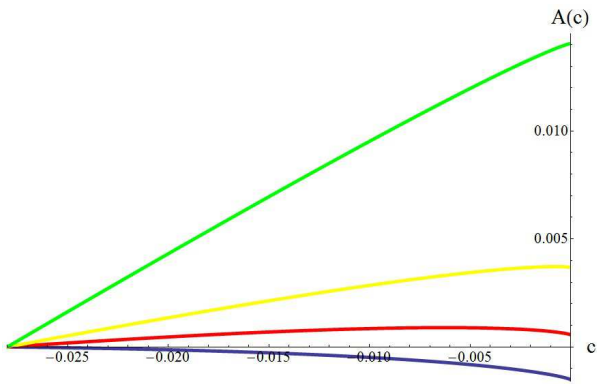}
\par\end{centering}
\caption{(Color Online) Plots $A(E)=-\frac{\alpha}{M_{s}}I_{i}^{E}(E)+\frac{J}{M_{s}}\mathbf{m}_{p}\cdot\mathbf{I}_{i}^{M}(E)$
as a function of energy $c=2E/\kappa_2$  over a range of spin torque current. From bottom to top:
 Blue = $0.77J_{c}$, Red = $1.08J_{c}$, Yellow = $1.55J_{c}$, Green = $3.10J_{c}$}
\label{fig-currents}
\end{figure}

Before going further though, it would be useful to consider how the
system acts in the absence of the noise. In such a case the
switching occurs when the energy current (1.34) is positive for all values
of energy between the starting position (say positive $z$ direction) and
the saddle point. Since the probability function $\rho_{i}^{E}(E,t)$
is always positive it stand to reason that a switch will only happen
if $0\leq-\alpha I_{i}^{E}(E)+J\mathbf{m}_{p}\cdot\mathbf{I}_{i}^{M}(E)$.
We plot this quantity as a function of energy for various values of
the spin-current $J$ in Fig. \ref{fig-currents}.  From this we also gain a useful reference value for
the critical current current which is $J_{c}=\frac{\alpha I_{i}^{E}(E_{sad})}{\mathbf{m}\cdot\mathbf{I}_{i}^{M}(E_{sad})}=0.00645M_{s}$.
The positive value signifies the tendency towards the switching. In the first example with $J=0.77J_c$
the noiseless system, being driven by the dissipation towards the stable position,  does not switch. It is worth noticing that in the presence of the noise the switching nevertheless does occur, but it takes exponentially long time. In the three other examples
$J\geq J_c$ and the magnetization current is always directed towards the saddle. Therefore even the noiseless system does switch and the noise serves to introduce an uncertainty in the switching time.

\begin{figure}
\begin{centering}
\includegraphics[width=8cm]{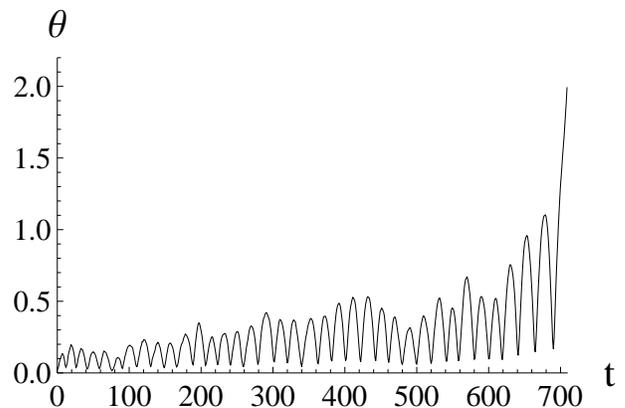}
\par\end{centering}
\caption{A typical realization of $\theta$ as a function of time (in units
of $(\gamma M_{s})^{-1}$) for $J=1.08J_{c}$.}
\label{fig-theta}
\end{figure}

Putting the thermal noise back into the system, we set the noise strength
parameter to $D_{th}=0.00001\gamma M_{s}$. Simulations are then
run by starting each particle at $\theta=0$, allowing it to come
into thermal equilibrium with the system, turning the current on and
calculating how long it takes for it to go past the saddle point into
the second well. This is done for many particles for a given current
value and over several different current values.
A typical trajectory of the system is represented by the graph
$\theta$ as a function of time in Fig. \ref{fig-theta} for $J=1.08J_{c}$. It may be seen that it takes many
revolutions before the system finally switches to the basin of attraction of the true stationary points at $t\approx 700$.

\begin{figure}
\includegraphics[width=4cm]{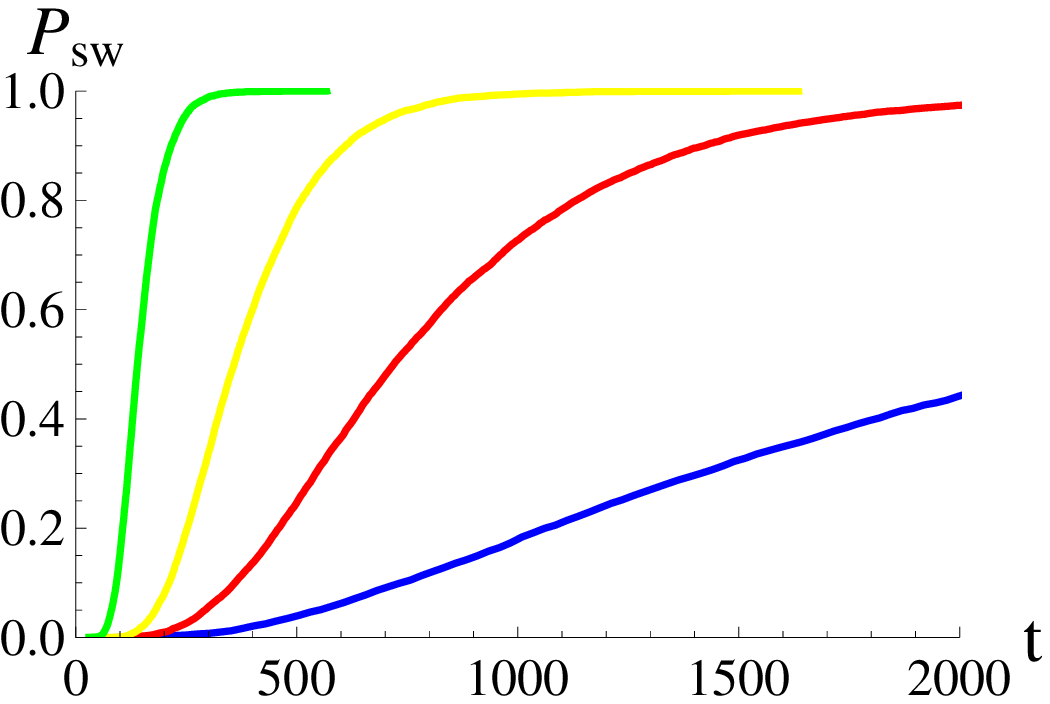}\includegraphics[width=4cm]{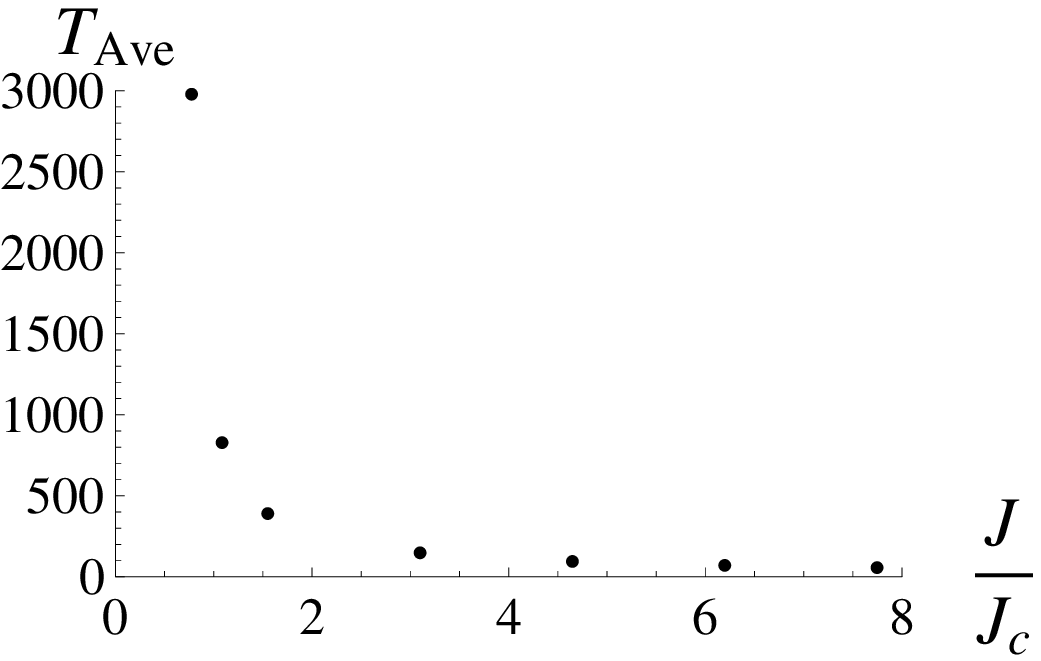}
\caption{(Color Online) (Left) Shows the switching probability as a function of time (in units
of $(\gamma M_{s})^{-1}$) for various current values. (From left to right: Green = $3.10J_{c}$,
Yellow = $1.55J_{c}$, Red = $1.08J_{c}$, Blue = $0.77J_{c}$). (Right)
Shows the average switching time (in units of $(\gamma M_{s})^{-1}$)
as a function of $\frac{J}{J_{c}}$.}
\label{fig-switch}
\end{figure}

Let us estimate the contribution of the nonequilibrium noise to the switching process under realistic experimental conditions. From Eq. (\ref{DI}) we obtain the relationship between equilibrium noise and non-equilibrium noise
\begin{equation}
\frac{D_{\mathrm{neq}}}{D_{\mathrm{eq}}}= 
\frac{\hbar^2\gamma}{8 e k_B M_s\mathcal{V}\alpha}\frac{I}{T}.  
\label{neqtoeq_I}
\end{equation}
Replacing $I$ with the critical current
\begin{equation}
I_c = J_c \frac{4M_s e\mathcal{V}}{\hbar}
\end{equation}
where $J_c$ is the minimum spin current needed to cause a switch in the absense of noise, 
we obtain 
\begin{equation}
\frac{D_{\mathrm{neq}}}{D_{\mathrm{eq}}}= 
\frac{\hbar\gamma}{2 k_B \alpha}\frac{J_c}{T}. 
\label{neqtoeq_I2}
\end{equation}
Using the material parameters in Ref. [\onlinecite{Katine00}], $M_{s}=1440$
emu, nanopillar volume = $1.97\times10^{-17}cm^{3}$, and switching
current $I_{c}\approx10^{9}A/cm^{2}$, the shot noise, $D_{0}$, at
the critical current is equivalent to the thermal noise, $D_{th}$, of
temperature $T_c\approx 15$ K. Even though the theoretical switching current
and the experimentally determined switching current of Ref. [\onlinecite{Katine00}] 
differ by more than an order of magnitude (such a discrepancy is mentioned in many experimental studies, including Ref. [\onlinecite{Katine00}])  
the theoretical approach 
can be used to gain insight into how the shot noise at the switching
current scales with the parameters of a  nanopillar.

If we set $D_{0}=D_{th}$ again and include our calculation of the
switching current we find $T_{c}\propto J_{c}/(\alpha P)$ where $J_{c}$
is the  critical spin-torque current needed to cause a spin-flipping
event (in the absence of noise) and $P$ is the degree of polarization of the current.
$J_{c}$ is determined by the relative values of $H_{k}$ and $M_{s}$
and has the units of magnetization. For the simple case with only easy-axis anisotropy 
($E=-\frac{1}{2}\mu_{0}H_{k}M_{s}\cos^{2}\theta$),
$J_{c}=\alpha H_{k}$, and therefore $T_{c}\propto\frac{H_{k}}{P}$.
Therefore for materials and/or geometry with a bigger anisotropy field the 
shot noise may be a dominant source of noise at temperatures well above $15K$. 
Moreover, in the simplest case where we only have uniaxial anisotropy and an applied external field along the easy axis it can be shown that
\begin{equation}
J_c=\alpha(H_k + H_{{\mathrm{ext}}}).
\end{equation}
This means the relative strength of the non-equilibrium noise scales with the potential well depth of our system.  By applying an external field we can increase the maximum allowed current before a switch takes place and thus increase the importance of non-equilibrium noise.

Since the initial condition is taken out of a stationary distribution (without the spin current) and
subsequent evolution is subject to the Langevin noise, the time of the switching is a random quantity.
The percentage of trial systems that have switched as a function of time is shown in the left panel of Fig. \ref{fig-switch}
for four different values of the spin-current. The time derivatives of these graphs provide probability distribution functions of
the switching time. One may then evaluate the first moment of these distributions which gives the mean switching time for a given
value of the spin current. The right panel of Fig. \ref{fig-switch} shows such a mean switching time as a function of $J/J_c$.
One may notice that for $J\leq J_{c}$ the switching time grows
exponentially, while for $J\geq J_{c}$ the switching time becomes
relatively short (although it is still substantially longer than the inverse precession frequency).

\section{Conclusion}
\label{sec-concl}

We would like to conclude with the following remarks.
The study of noise in dynamical magnetic systems is
a broad and fascinating field. In particular, in view
of potential applications of magnetic nanodevices,
both equilibrium and nonequilibrium noise
may
play an important role.
For example, the stability of magnetic storage devices is
strongly influenced by thermal fluctuations.
The functionality of new generation technologies (such as the magnetic random
access memory (MRAM) \cite{Prinz}
with STT writing
or the racetrack memory \cite{ParkinRacetrack})
is largely based on the spin torque phenomenon. The
latter is a nonequilibrium effect and thus besides
the temperature also nonequilibrium sources of noise
may play an important role.

A way to introduce fluctuations into magnetization dynamics
is to add a random component
to the effective field or to the current in the
phenomenological LLG equation.
The noise is then defined by the value of its
correlator (and its higher order cumulants).
The determination of the noise correlator
is of great importance as it defines the noise
properties. In addition it may
give insight into the physical context.

A powerful and very  flexible  tool is the Langevin approach
based on the Keldysh path integral formalism. Starting
from a microscopic model one derives the equations of
motion for the magnetic system. Fluctuations naturally arise
as a generic feature of the Keldysh approach. We have demonstrated
the applicability of this method to magnetic systems
on the example of spin shot noise in magnetic tunnel
junctions. The spin shot noise correlator arose naturally, as a
consequence of the sequential tunneling approximation,
in second order in spin flip processes.

The Keldysh formalism is however not restricted to the system
described above. In particular it may be used in the
context of nonuniform magnetic textures
(as domain walls for instance). Promising advances in
this direction have already been reported \cite{DuineLucassen08}
and demonstrate the versatility of the method
as well as inspire us with curiosity
about future developments.

Finally, to investigate the influence of the spin shot noise
on spin torque switching rates we have generalized
the Fokker-Planck approach of
Ref[\onlinecite{ApalkovPRB}]. We have shown that
the nonequilibrium noise manifests itself in a renormalized
effective temperature. In particular at low temperatures
we could observe a significant variation of the noise
with orbit energy, reflecting "cold" and "hot" trajectories
of the magnetization vector with respect to the noise intensity.

\section*{Acknowledgements}
J. Swiebodzinski and A. Chudnovskiy  acknowledge financial support from DFG through
Sonderforschungsbereich 508 and T. Dunn and  A. Kamenev acknowledge support from  NSF
grant DMR-0804266.

\bibliographystyle{unsrt}
\bibliography{jacek_dr}

\end{document}